# AlGaN/GaN asymmetric graded-index separate confinement heterostructures designed for electron-beam pumped UV lasers

SERGI CUESTA,[1,*] YOANN CURÉ,[1] FABRICE DONATINI,[2] LOU DENAIX,[1] EDITH BELLET-AMALRIC,[1] CATHERINE BOUGEROL,[2] VINCENT GRENIER,[1] QUANG-MINH THAI,[3] GILLES NOGUES,[2] STEPHEN T. PURCELL,[3] LE SI DANG,[2] AND EVA MONROY[1]

[1]*Univ. Grenoble-Alpes, CEA, Grenoble INP, IRIG, PHELIQS, F-38000 Grenoble, France*
[2]*Univ. Grenoble-Alpes, CNRS, Institut Néel, F-38000 Grenoble, France*
[3]*Institut Lumière Matière, CNRS, Univ. Lyon, Univ. Claude Bernard Lyon 1, F-69622 Villeurbanne, France*
*\*sergi.cuestaarcos@cea.fr*

**Abstract:** We present a study of undoped AlGaN/GaN separate confinement heterostructures designed to operate as electron beam pumped ultraviolet lasers. We discuss the effect of spontaneous and piezoelectric polarization on carrier diffusion, comparing the results of cathodoluminescence with electronic simulations of the band structure and Monte Carlo calculations of the electron trajectories. Carrier collection is significantly improved using an asymmetric graded-index separate confinement heterostructure (GRINSCH). The graded layers avoid potential barriers induced by polarization differences in the heterostructure and serve as strain transition buffers which reduce the mosaicity of the active region and the linewidth of spontaneous emission.

## 1. Introduction

There is a high demand of ultraviolet (UV) lasers for applications in the fields of medicine and biotechnology [1], as well as in 3D printing [2] and non-line-of-sight communication [3]. Presently, this spectral range is covered by gas lasers (ArF, KrF, XeF) or lasers based on frequency conversion (Nd:YAG). III-nitride semiconductor laser diodes are promising candidates to provide an efficient semiconductor-based alternative [4,5]. However, current injection is a major problem for implementing AlGaN-based laser diodes. The p-type doping of AlGaN is particularly challenging [6], and the high doping concentrations required to achieve p-type conductivity degrade the material quality and increase absorption losses. This is one of the main reasons why laser diodes emitting below 370 nm show a dramatic increase in threshold current density [7–11].

The implementation of optically-pumped semiconductor lasers is a possible solution to the carrier injection problem. Compared to current injection laser diodes, optically-pumped laser structures require a much simpler heterostructure design. The heterostructure must provide electrical and optical confinement, with the active media as close to the surface as possible to ensure the collection of photogenerated carriers, and doping is no longer required. Some experimental realizations of optically-pumped AlGaN UV lasers can be found in the literature [11–13].

Another alternative is to pump the semiconductor structure with a high energy (5-20 keV) electron beam. Its advantages include higher flexibility in the choice of materials for the active medium due to the absence of doping or electrical contacts, as well as higher radiative recombination efficiency since the electrons and holes generated by impact ionization share the same distribution in the active medium. Compared with optical pumping, electron-beam pumping offers the possibility to make devices more compact, and the ability of the electron beam to penetrate deeper into the structure allows greater freedom in the design of the optical cavity avoiding surface-related loses. Furthermore, electron beam pumping is adapted to a large spectral range, whereas in the case of optical pumping, the wavelength of the output laser must be longer than the wavelength of the pumping laser.

The electron-beam pumping technology has allowed the fabrication of ZnSe-based pulsed lasers that emit up to 600 W at 535 nm [14]. There are some studies of electron beam pumped AlGaN for the implementation of UV lamps [15–22]. Wunderer *et al*. demonstrated UV lasing from an AlGaN/GaN heterostructure containing InGaN quantum wells (QWs) under pulsed electron beam excitation at 77 K [23]. The emission peaked at 384 nm, with a threshold of 174 µA at an acceleration voltage of 18 kV (≈160 kW/cm$^2$ assuming the beam focused to a spot diameter of 50 µm). At the same time, Hayashi *et al*. showed lasing from an In-free AlGaN/GaN structure under pulsed electron beam excitation at 107 K. A separate confinement heterostructure (SCH) containing a 10-period GaN/AlGaN multi-quantum well (MQW) displayed stimulated emission around 353 nm with a threshold power density around 230 kW/cm$^2$ [24]. These demonstrations have opened a new line of research, but new device architectures are required with the prospect of achieving room temperature lasers.

Here, we present a study of undoped AlGaN/GaN SCHs designed to operate under electron beam injection with an acceleration voltage ≤ 10 kV. We discuss the effect of spontaneous and piezoelectric polarization on the carrier diffusion, and demonstrate that the performance is improved using an asymmetric graded-index separate-confinement heterostructure (GRINSCH).

## 2. Design

Figure 1(a) presents a schematic description of the laser structures under study. The choice of bulk GaN as a substrate aims at minimizing the dislocation density. The active region consists of a 10-period GaN/AlGaN MQW inserted in an AlGaN/AlGaN SCH, where the core (bottom/top inner cladding, BIC/TIC in the figure) has lower Al mole fraction than the outer claddings (bottom/top outer cladding, BOC/TOC in the figure). The waveguide is induced by the variation in the refraction index between the cladding layers. In some samples, the alloy transition between the AlGaN cladding layers is gradually switched in a so-called graded layer, with the ternary alloy composition varying linearly along the growth axis.

The design of the structures was carried out taking into account both optical and electrical considerations: (i) the bottom cladding layer must be thick enough to prevent waveguide losses through the absorbing GaN substrate, (ii) the penetration depth of the electron beam in the structure imposes a limit to the thickness of the upper layers, and (iii) the MQW must emit at the targeted wavelength. Therefore, on the one hand, we studied the confinement of the optical mode in the waveguide using a commercial finite-element analysis software (Comsol Multiphysics) with the values of refractive indices extracted from [25,26]. On the other hand, we conducted Monte Carlo simulations using the CASINO software to predict the path of the electron beam with a targeted operation voltage ≤ 10 kV. Finally, for the design of the MQW geometry, the electronic band profiles were modelled in one-dimension using the Nextnano$^3$ 8-band k·p self-consistent Schrödinger-Poisson solver [27], with the material parameters described in [28]. The structures were considered pseudomorphic on the GaN substrate, and the spontaneous and piezoelectric polarization, as well as the band structure deformation potentials, were taken into account. The thicknesses of QWs and barriers were tuned to attain the targeted operation wavelength. The simulations were performed incorporating an electron density of $10^{18}$ cm$^{-3}$ to mimic operation conditions.

The final architectures, summarized in Table 1, were obtained after several iterations of optical and electronic simulations. As an example, Fig. 1(b) presents the simulation of the first optical mode confined in the SCH for sample S1, and Fig. 1(c) shows the predicted trajectory of the electrons in the same sample. The calculations show that the vast majority of carriers are absorbed in the active region for an acceleration voltage of 7.5 kV. In comparison with the design of Hayashi *et al.* [24], here the thickness of the upper layers were reduced to operate at a lower acceleration voltage (7-10 kV instead of 15 kV), and the waveguide core was also smaller, to improve the optical confinement. Thus, S1 presents an optical confinement factor of 4.1%, which is an improvement with respect to the ≈ 6% reported by Hayashi *et al.* [24] for 10 × 3 nm QWs, since we use much narrower QWs (10 × 1.4 nm). Note that this QW width is chosen to attain 355 nm emission at room temperature. Using the Varshni's equation that describes the variation of the band gap with temperature for GaN [29], our QWs should present an emission energy around 48 meV higher with respect to the structure of Hayashi *et al*, who attained 355 emission nm at 107 K. The spectral shift is achieved here by increasing the Al content of the AlGaN layers and reducing thickness of the active QWs.

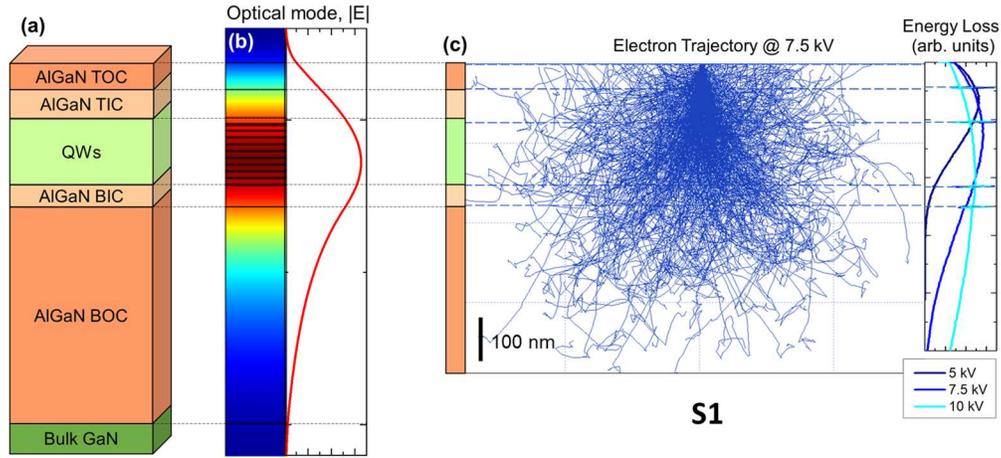

Fig. 1. (a) Schematic view of the laser structure. The following layers are grown sequentially on top of GaN substrate: bottom outer cladding (BOC), bottom inner cladding (BIC), MQWs active region, top inner cladding (TIC) and top outer cladding (TOC). (b) Simulation of the first optical mode distribution through the SCH for sample S1. (c) Simulation of the electron beam trajectory in the structure S1. The central panel shows the trajectory of the electrons generated by an electron beam under 7.5 kV. The panel on the right side shows the electron energy loss profile for various acceleration voltages.

Table 1. Description of samples under study (thickness and Al content of the layers, following the general design in Fig. 1(a)), their optical confinement factor (OCF), and experimental measurement of the full width at half maximum of the ω-scan of the (0002) X-ray reflection of the MQW (Δω).

|  | S1 | S2 | S3 |
|---|---|---|---|
| TOC | 44.4 nm $Al_{0.2}Ga_{0.8}N$ | 26.4 nm $Al_{0.2}Ga_{0.8}N$ | 44.2 nm $Al_{0.3}Ga_{0.7}N$ |
| Graded |  | 35.2 nm | 48.7 nm |
| TIC | 59.5 nm $Al_{0.1}Ga_{0.9}N$ | 41.4 nm $Al_{0.1}Ga_{0.9}N$ |  |
| QWs | 10×(1.4 nm GaN /9.8 nm $Al_{0.1}Ga_{0.9}N$) | 10×(1.3 nm GaN /9.7 nm $Al_{0.1}Ga_{0.9}N$) | 10×(1.3 nm GaN /9.7 nm $Al_{0.1}Ga_{0.9}N$) |
| BIC | 35.5 nm $Al_{0.1}Ga_{0.9}N$ | 17.6 nm $Al_{0.1}Ga_{0.9}N$ | 17.7 nm $Al_{0.1}Ga_{0.9}N$ |
| Graded |  | 35.2 nm | 35.4 nm |
| BOC | 355 nm $Al_{0.2}Ga_{0.8}N$ | 332 nm $Al_{0.2}Ga_{0.8}N$ | 334 nm $Al_{0.2}Ga_{0.8}N$ |
| OCF | 4.1% | 3.7% | 3.8% |
| Δω | 208 arcsec | 128 arcsec | 158 arcsec |

## 3. Experimental

The laser structures described in Table 1 were synthesized on free-standing, non-intentionally doped, n-type conductive GaN(0001) substrates using plasma-assisted molecular-beam epitaxy, with the growth conditions described elsewhere [30]. The substrate temperature was $T_S$ = 720 ºC and the growth rate was 0.5 monolayers per second (ML/s). Note that in wurtzite GaN and AlGaN, the thickness of 1 ML is approximately 0.25 nm.

The periodicity of the MQW structures was analyzed by high resolution X-ray diffraction (HRXRD) in a Rigaku SmartLab diffractometer. Additional structural studies were conducted on a thin lamella prepared by focused ion beam, using high-angle annular dark-field (HAADF) scanning transmission electron microscopy (STEM) performed on a FEI Thémis microscope operated at 200 kV.

Cathodoluminescence (CL) experiments were carried out in a FEI Inspect F50 field-emission SEM equipped with an iHR550 spectrometer fitted with an Andor Technology Newton DU940 BU2 spectroscopic charge-coupled device (CCD) camera. The beam spot diameter was ~10 nm on the focal point, the scanning area was a square of 4 μm$^2$, the accelerating voltage was varied from 5 to 20 kV and the injection current was around 100 to 150 pA. The CL was collected from the top of the as-grown sample at room temperature.

A set of laser bars with a resonator length of 1 mm were prepared by mechanical cleaving along the (10-10) *m*-plane of GaN. The facets were uncoated. Photoluminescence (PL) measurements were obtained under pulsed excitation using a Nd-YAG laser (λ = 266 nm, pulse duration = 0.5 ns, repetition rate = 8 kHz) at room temperature. Cylindrical lenses were inserted to shape the laser beam into a 100-μm-wide stripe, and to control the beam uniformity (variation of intensity <10% along the cavity). All tests were carried out at room temperature. The PL edge emission was collected by a Jobin Yvon HR460 monochromator equipped with a UV-enhanced CCD camera. A ThorLabs WP25M-UB UV polarizer was used for the polarization-dependent measurements.

## 4. Results and discussion

In a first step, the geometry and structural quality of the samples was validated by HRXRD and STEM. Figure 2(a) shows HRXRD θ–2θ scans around the (0002) reflection of GaN for samples S1, S2 and S3, compared with a theoretical calculation using the GlobalFit2 software from Rigaku Co. The calculations used S1 design described in Table 1, and the structure was assumed pseudomorphic on the BOC layer, which was allowed to relax partially (relaxation in the range of 50%). The more intense peak of the diffractogram is attributed to the GaN substrate, and the second highest, located at slightly higher angles, to the TOC and BOC layers. The reflection of the MQW presents several satellites, which allow a precise determination of the MQW period. To compare the mosaicity of the samples, the full width at half maximum (FWHM) of the ω-scan around the (0002) X-ray reflection of the MQW is presented in Table 1 (Δω) for all the samples. HAADF-STEM images for sample S1 are shown in Fig. 2(b), where clear contrast corresponds to layers with higher Ga concentration. The zoomed image on the right side reveals a QW thickness of 1.4 nm.

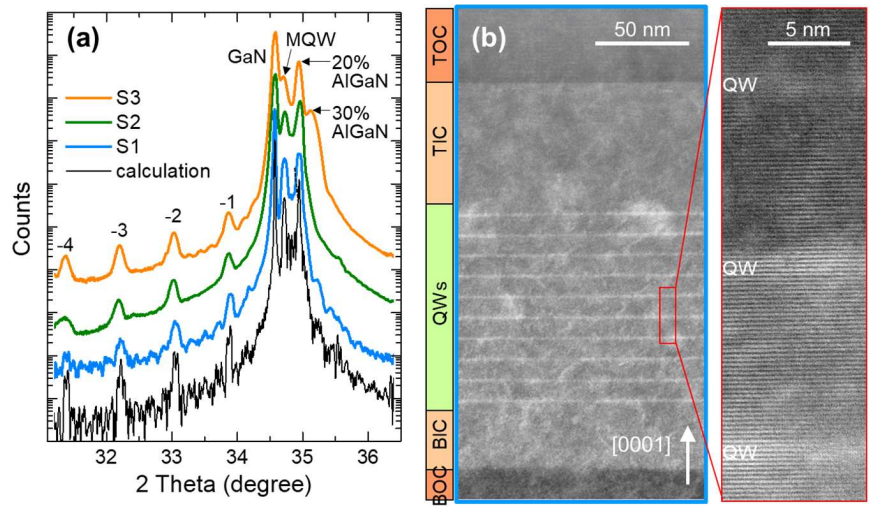

Fig. 2. (a) HRXRD θ-2θ scan around the (0002) reflection of GaN for samples S1 S2 and S3. Labels identify the reflections originating from the GaN substrate, the $Al_{0.2}Ga_{0.8}N$ and $Al_{0.3}Ga_{0.7}N$ layers and the MQWs with several satellites. Reflections from $Al_{0.1}Ga_{0.9}N$ layers are not resolved due to their proximity to the QW reflection. Experimental results are compared with a theoretical calculations of the diffractogram for S1. (b) HAADF-STEM image of the inner section of the heterostructure in sample S1. On the right side, zoomed image showing 3 QWs.

The optical characterization of sample S1 is summarized in Fig. 3. Figure 3(a) shows the variation of the CL spectra as a function of the acceleration voltage, $V_A$. Increasing $V_A$ implies changing the penetration depth of the electron beam, as illustrated in the right panel of Fig. 1(c). Electron-hole pairs generated by impact ionization deep inside the structure will have to diffuse towards the MQW to emit at the desired wavelength. Therefore, the evolution of the CL as a function of $V_A$ gives information about the efficiency of the carrier collection process. This measurement is performed with the injection current remaining approximately constant. The CL spectra are normalized by dividing by the acceleration voltage and injection current, so that their comparison provides a direct view of the evolution of the conversion efficiency. Looking at the figure, at low acceleration voltage, the emission is dominated by recombination in the MQW (at 354 nm). This means that carriers generated close to the surface are efficiently transferred to the MQW. However, a very weak signal around 322 nm is visible for $V_A = 5$ kV, and it is assigned to recombination in the TOC layer. Nevertheless, in view of the intensity of the curves, the efficiency of the device is maximum in the range of $V_A = 5$-7 kV. At higher acceleration voltages, a secondary peak assigned to $Al_{0.1}Ga_{0.9}N$ (inner cladding, at 342 nm) appears, and it becomes dominant for $V_A > 12$ kV. This behavior points out that carriers generated in the BIC layer have problems to diffuse to the MQW, and they recombine before reaching the wells.

As explained in the experimental section, the CL spectra presented here were recorded using a CL-SEM system, which does not reach high enough current densities to attain the amplified spontaneous emission regime. Furthermore, the experiments were performed in as-grown material, without mirror processing, so that lasing effect is not expected.

To explore the lasing threshold, laser bars with a cavity length of 1 mm were optically pumped using the 266 nm Nd-YAG laser source at room temperature. Figure 3(b) displays the room-temperature PL emission spectra at various pumping power densities. Laser emission appears at 355 nm above the threshold power density of 210 kW/cm$^2$. Figure 3(c) presents the variation of the PL intensity and its FWHM as a function of the pumping power density, showing an abrupt superlinear increase of the emission at the threshold together with a drop of the FWHM by several orders of magnitude (note that the measured values are limited by the resolution of the setup, around 0.1 nm). Figure 3(d) shows the polarization properties of the stimulated and spontaneous emission from sample S1. The stimulated emission display a clear dependence on the polarization angle, with a transverse-electric : transverse-magnetic output power ratio TE:TM = >30:1.

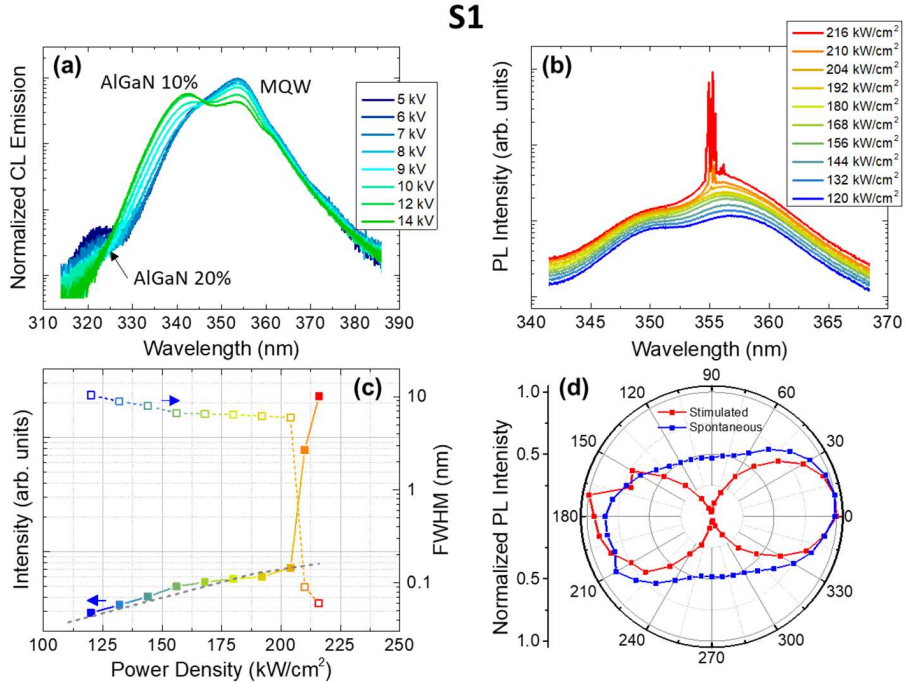

Fig. 3. (a) CL spectra of sample S1 at different acceleration voltages. The CL spectra are normalized by dividing by the acceleration voltage and injection current. (b) Photoluminescence spectra in semi-logarithmic scale of sample S1 (cavity length = 1mm) at different pumping power. (c) Emitted photoluminescence intensity (full markers) and FWHM (empty markers) in function of the excitation power density. The grey dashed line represents a linear tendency. (d) PL intensity collected as a function of the polarization angle (0 degrees corresponds to TE polarization).

As discussed above, the intensity of the CL emission at 342 nm in Fig. 3(a) reveals carrier transfer losses from the BIC layer to the QWs. The origin of the problem becomes evident when looking at the band diagram of S1, depicted in Fig. 4(a). The difference in spontaneous and piezoelectric polarization between the waveguide core and the outer cladding results in a strong band bending that favors electron accumulation at the TIC/TOC interface and hole accumulation at the BOC/BIC interface (red arrows in the figure). The perturbation of the bands hinders carrier transfer to the MQW. This leads us to consider the implementation of a graded index of refraction separate confinement heterostructure (GRINSCH), which promotes the diffusion of electrons towards the wells while the waveguiding properties of the structure are preserved [31,32]. The capabilities of the GRINSCH design are generally not used in III-nitride laser diodes since they introduce a perturbation of the doping distribution due to the variation of polarization in the graded layers [33,34]. However, these effects are not relevant in an electron-beam pumped device so that the advantages of graded index layers can be fully exploited. Therefore, the design for S2 (see Fig. 5(a)) is a modification of S1 where graded interfaces are inserted between the inner and outer cladding layers, resulting in the band diagram in Fig. 4(b), with smooth energy transitions at the heterointerfaces (arrows in the figure).

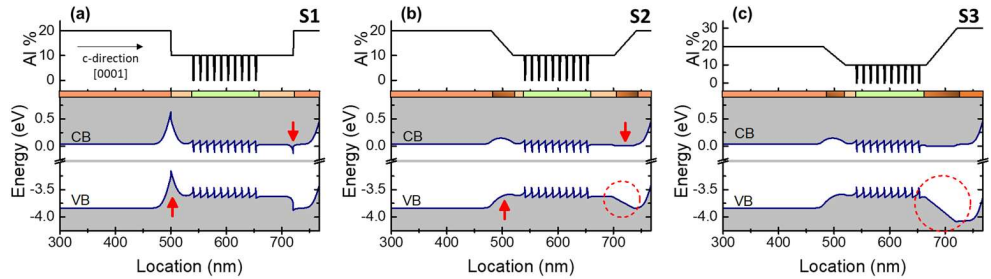

Figure 4. Alloy concentration along the growth direction and simulated band diagram for the conduction and the valence band for (a) S1, (b) S2 and (c) S3. Red marks aid to identify the main changes in the bands from one structure to another.

Experimentally, the effect of the GRINSCH is directly noticeable in the CL spectra of S3, shown in Fig. 5(b). Note that here, the CL spectra are also normalized by dividing by the acceleration voltage and injection current, so that their comparison can provide a direct view of the evolution of the efficiency. The spectra present a main peak at 352 nm assigned to the QWs and a secondary peak at 339 nm attributed to the $Al_{0.1}Ga_{0.9}N$ inner cladding layers. A third peak appearing at 324 nm under high acceleration voltage ($V_A > 10$ kV) is assigned to the emission of the $Al_{0.2}Ga_{0.8}N$ BOC layer when the energy of the injected electrons is high enough to penetrate deeper than the GRINSCH region [see the penetration depth represented in the right panel of Fig. 1(a)]. At the same time, the integrated emission intensity starts to decrease, probably due to reabsorption of the luminescence of the BOC by the upper layers. In comparison with the CL data from S1, in Fig. 2(a), the intensity of the emission from $Al_{0.1}Ga_{0.9}N$ for $V_A > 10$ kV has considerably decreased when compared with the MQW line. This confirms an improvement of the transfer from the BIC to the MQW. However, the line at 339 is clearly resolved at low accelerating voltages, which implies that carriers recombine in the TIC layer without diffusing to the MQW.

We also noticed that the emission peaks from S2 are significantly narrower than those from S1, although both samples have the same alloy compositions, approximately the same QW thickness and they emit at the same wavelengths. The reduction of the FWHM can be attributed to a lower defect density in the structure with GRINSCH due to the smooth change in the lattice parameter between cladding layers. The comparison of the FWHM of the X-ray ω-scan around the (0002) reflection of the MQWs from both samples (Δω in Table 1) corroborates the improvement in crystalline quality, since Δω decreases from 208 arcsec for S1 to 128 arcsec in S2. This result is consistent with the structural studies of Sun *et al*. [31], who showed that such compositionally graded AlGaN layers may serve as a strain transition buffer which blocks threading defects.

The room-temperature PL spectra presented in Fig. 5(c) shows stimulated emission at λ = 358 nm with a power density threshold of 180 kW/cm$^2$. The strong improvement in CL leads to a moderate improvement of the threshold. This is due to the fact that under optical pumping the excited region is limited to the topmost ≈ 100 nm, i.e. the excitation is much more superficial than in the case of electron pumping. Therefore, the improvement is expected to be larger in the case of electron pumping. Figure 5(d) illustrates the polarization of the emitted light. The polarization of the stimulated emission is clearly defined in the TE mode direction, and it vanishes in the perpendicular TM mode direction (TE:TM = >30:1). In contrast, spontaneous emission is not polarized in any particular direction, as expected.

In summary, the optical characterization data of S2 represents a significant improvement with respect to S1. However, further efforts are needed to reduce the carrier recombination rate in the top cladding layers. To address this issue, we consider the implementation of an asymmetric GRINSCH in sample S3, with the design illustrated in Fig. 5(e). The top graded layer extends now to the edge of the MQW and the Al mole fraction of the top outer cladding goes up to 30%. The purpose of these modifications, outlined with dashed red circled in Figs. 4(b) and 4(c), is to force carriers generated in the top layers to diffuse towards the active region. Following this design, the CL spectra of sample S3 in Fig 5(f) shows the predominance of the MQW peak at 351 nm. A peak at 323 nm, assigned to the $Al_{0.2}Ga_{0.8}N$ BOC, appears only at high acceleration voltage, when the carriers penetrate deeper than the collection region defined by the GRINSCH. For operation at $V_A ≤ 7$ kV, 99% of the emission stems from the MQW. For comparison purposes, in the case of S2 at the same acceleration voltage, the light emitted from the MQW was 75% of the total emission. The PL spectra of S3 in Fig. 5(g) shows stimulated emission at λ = 353 nm with a room-temperature threshold power density of 180 kW/cm$^2$. The polarization feature in Fig. 5(h) presents the same behavior as S2, with the stimulated emission clearly polarized in TE direction (TE:TM = >30:1).

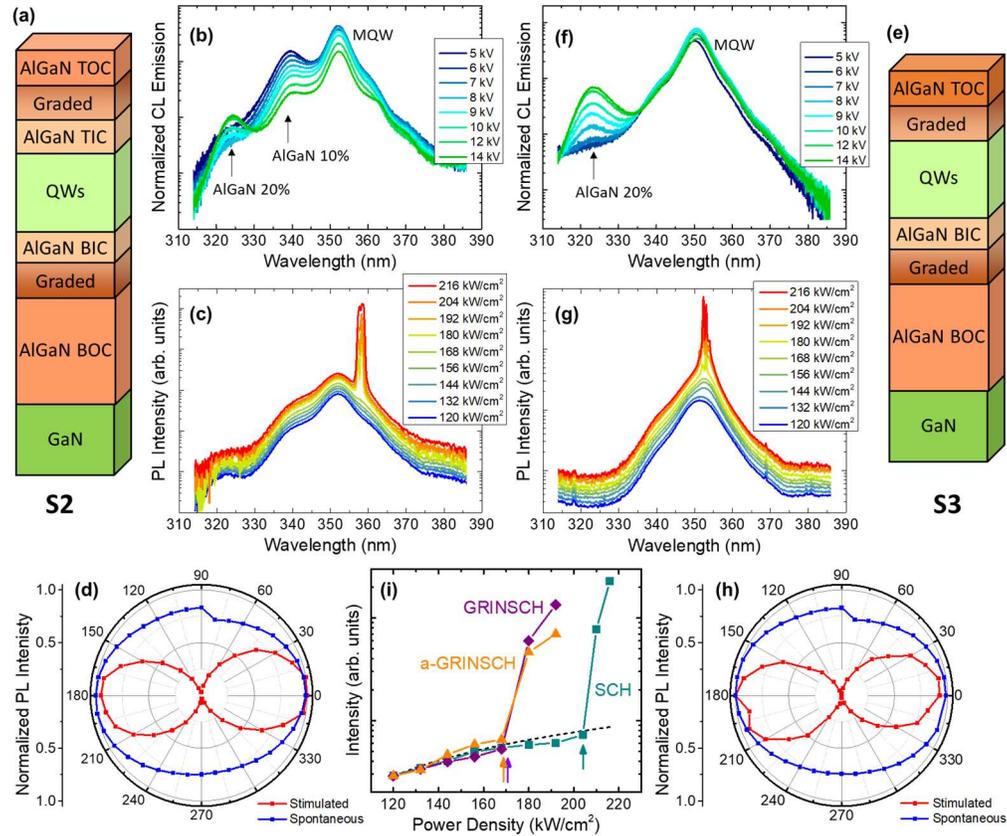

Fig. 5. (a) Schematic view of the laser structure S2. (b) CL spectra recorded at different acceleration voltages (S2). The CL spectra are normalized by dividing by the acceleration voltage and injection current. (c) PL spectra in semi-logarithmic scale at different pumping power densities for a 1-mm-long cavity (S2). (d) Normalized PL intensity as a function function of the polarization angle (S2). (e) Schematic view of the laser structure S3. (f) CL spectra at different acceleration voltages (S3). The CL spectra are normalized by dividing by the acceleration voltage and injection current. (g) PL spectra in semi-logarithmic scale at different pumping power densities for a 1-mm-long cavity (S3). (h) Normalized PL intensity as a function of the polarization angle (S3). (i) PL peak intensity as a function of the excitation power density for samples S1 (SCH design), S2 (GRINSCH) and S3 (asymmetric GRINSCH). The dark dashed line represents a linear tendency.

A comparison of the optical performance in terms of laser threshold is presented in Fig. 5(i), where we summarize the data from the three samples under study. The GRINSCH implementation leads to an improvement of the lasing performance in comparison with the initial SCH design, reducing by 15% the threshold power density. The asymmetric GRINSCH follows the same tendency in addition to improving even further the carrier collection efficiency under electron beam pumping.

It should be noted that the optically-pumped threshold power densities presented here could be decreased by eliminating the top cladding layers. However, this would result in a degradation of the performance under electron pumping, due to the different carrier distribution between an optically-pumped and an electron-beam pumped device. The measurement of the optical threshold for the different SCH designs provides comparative information about the carrier transfer from the top layers to the active region and the optical loses in the waveguides. Our experiments demonstrate that the implementation of a GRINSCH results in a relevant improvement of the collection of carriers generated by impact-ionization while maintaining laser emission at 353 nm with a threshold power density lower than that of the equivalent SCH design.

## 5. Conclusion

In this work, we presented AlGaN/GaN heterostructures designed to operate under electron beam injection for acceleration voltages ≤ 10 kV, and displaying laser emission at room temperature under optical pumping. The implementation of a GRINSCH improves the collection of carriers towards the active region without any negative impact in the quality of the waveguide. Furthermore, the smooth transition between ternary alloys in the compositionally graded layers results in an improved structural quality of the active MQW, thus reducing the threshold compared with the SCH. In addition to that, an asymmetric design of the GRINSCH is introduced, to further improve the transfer of carriers towards the active region. The combination of these modifications could lead to new electron-pumped laser devices with high performance in the UV emission range.

## Funding

This work is supported by the French National Research Agency (ANR) via the UVLASE program (ANR-18-CE24-0014), and by the Auvergne-Rhône-Alpes region (grant PEAPLE).

## Acknowledgements

We benefited from the access to the technological platform NanoCarac of CEA-Minatech Grenoble in collaboration with the IRIG-LEMMA group.

## Disclosures

The authors declare no conflict of interest.